# Migration studies with a Compositional Data approach: a case study of population structure in the Capital Region of Denmark

**Authors:** Javier Elío[1], Marina Georgati[2], Henning S. Hansen[3], Carsten Keßler[4]

**Corresponding author**: Carsten Keßler

**Abstract**:

Data normalization for removing the influence of population density in Population Geography is a common procedure that may come with an unperceived risk. In this regard, data are constrained to a constant sum and they are therefore not independent observations, a fundamental requirement for applying standard multivariate statistical tools. Compositional Data (CoDa) techniques were developed to solve the issues that the standard statistical tools have with close data (i.e., spurious correlations, predictions outside the range, and sub-compositional incoherence) but they are still not commonly used in the field. Hence, we present in this article a case study where we analyse at parish level the spatial distribution of Danes, Western migrants and non-Western migrants in the Capital region of Denmark. By applying CoDa techniques, we have been able to identify the spatial population segregation in the area and we have recognized some patterns that can be used for interpreting housing prices variations. Our exercise is a basic example of the potential of CoDa techniques, which generate more robust and reliable results than standard statistical procedures, but it can be generalized to other population datasets with more complex structures.

**Keywords**: Population geography, migration, compositional data, cluster analysis, regression models

[1] Department of Planning, Aalborg University, Copenhagen, Denmark; ORCID: 0000-0003-0624-2345; javierdem@plan.aau.dk - javiereliomedina@gmail.com

[2] Department of Planning, Aalborg University, Copenhagen, Denmark; ORCID: 0000-0001-7794-8308; marinag@plan.aau.dk

[3] Department of Planning, Aalborg University, Copenhagen, Denmark; ORCID: 0000-0001-7004-0698; hsh@plan.aau.dk

[4] Department of Geodesy, Bochum University of Applied Sciences, Bochum, Germany; and Department of Planning, Aalborg University, Copenhagen, Denmark; ORCID: 0000-0002-9724-820X; carsten.kessler@hs-bochum.de

## 1. Introduction

In population geography, it is often more interesting to analyse proportions than absolute population numbers, such as the percentage of people in a region with an income below the poverty line, or the proportion of people fully vaccinated against COVID-19. The variable of interest is normalized so that it does not depend on the total population of the region (Dailey, 2006), thus allowing for an intuitive comparison of e.g. poverty or vaccinate rates across regions. This normalization, however, comes with an unperceived risk. The data are constrained to a constant sum (e.g., 1 for proportions or 100 for percentages) and, therefore, they are not independent: if the share of one subgroup increases, another one has to decrease to retain the sum (Pawlowsky-Glahn and Egozcue, 2006). Hence, standard statistical procedures may lead to spurious correlations, predictions outside the range, or have problems with sub-compositional coherence (Pawlowsky-Glahn, Egozcue and Tolosana-Delgado, 2011; Lloyd, Pawlowsky-Glahn and Egozcue, 2012).

These data are known as Compositional data (CoDa), which "*consist of vectors whose components are the proportion or percentages of some whole*" (Aitchison, 2002) – a very common kind of data that can be found in my kinds of analysis across a wide range of disciplines (Aitchison, 2002). Aitchison first described the theoretical background to handle such data based on log-ratio transformations (Aitchison, 1981, 1982); however, this approach is still not widely used despite all the associated issues that can arise when dealing with them. In fact, the compositional data methods have been mainly applied in the geosciences but even in this field it is not a standard procedure (Buccianti and Grunsky, 2014). For example, CoDa has been applied in soil and geochemical surveys (Zhang *et al.*, 2013; Grunsky, Mueller and Corrigan, 2014; Tepanosyan *et al.*, 2020; Zheng *et al.*, 2021), in water and groundwater studies (Buccianti, 2018), or for evaluating the link between indoor radon and topsoil geochemistry (Ferreira *et al.*, 2018). Outside of the geosciences, the technique is starting to be used across different studies in various fields, including the evaluation of urban water distribution (Ebrahimi *et al.*, 2021), health studies (McKinley *et al.*, 2020; Dumuid *et al.*, 2021; Verswijveren *et al.*, 2021), nutrition research (Corrêa Leite, 2019), and to forecast energy consumption structures (Wei *et al.*, 2021).

Geography is no exception regarding the lack of use of CoDa techniques and it is common that geographers apply standard statistical and geostatistical tools that have been designed for unconstrained data to analyse compositional data (e.g. percentage of young, working age and elderly population in a region; percentage of rented/owner households; unemployment rates). Lloyd, among others (Lloyd, 2010; Lloyd, Pawlowsky-Glahn and Egozcue, 2012), already warned about these problems and showed the tools for dealing with compositional data in population studies. Nonetheless, it seems the research community has not adopted these tools and only recently, they are starting to utilize CoDa techniques. For example, for evaluating social-spatial segregation (Cruz-Sandoval, Roca and Ortego, 2020), studying child mortality levels and trends (Ezbakhe and Pérez Foguet, 2020), forecasting population age structure (Wei *et al.*, 2019), or visualizing three part compositions in demographic analysis (Schöley, 2021). For migration studies, to the best of our knowledge, there are no studies that use the full range of CoDa techniques to analyse migration data, and only (Nowok, 2020) proposed the use of ternary diagrams for evaluating migration flows.

In this article, we therefore show the applicability of compositional data techniques in population geography, analysing the spatial population structure of the capital region of Denmark as a case study. Our aim is to stress the need of CoDa techniques in this type of studies to get robust and reliable results. We have used parish-level data for the year 2020 from Statistics Denmark to analyse the spatial distribution of the three main migrant categories defined by the national statistics: people of Danish origin, Western migrants and non-Western migrants. After performing a log-ratio transformation (i.e., balances), we carried out a hieratical cluster analysis for detecting areas where migrants settle down preferentially in the Capital region of Denmark. Furthermore, we explore the applicability of balances for evaluating the association between migration (and type of migrants) and housing prices through a linear model between median house prices and population structure.

It is well known that house prices and migration are closely related with profound implications on urban planning. There is a two-way causal relationship between migration and house prices (e.g. (Jeanty, Partridge and Irwin, 2010; Lin *et al.*, 2018)). On the one hand, a rise in house prices will increase a household's housing equity and, therefore, ability to migrate, since homeowners have a higher financial flexibility for purchasing a new house. At the same time, high house prices can make the house unaffordable, thus limiting the number of potential buyers. This way, price differences between the region where migrants live and regions where they intend to move affect in- and out- migration rates. Moreover, the expectation of future house prices also plays an important role in the decision to move (Peng and Tsai, 2019). On the other hand, migration increases the housing demand and, as a consequence, the prices (Wang, Hui and Sun, 2017). An example of this effect has been found in Sweden where "*a 1% increase in the foreign-born population results in a 0.8% increase in house prices, which increases to 1.2% if internal migration is also accounted for*" (Tyrcha and Abreu, 2019). However, they used in their models data that contain relative information (e.g., percentage of young, working age and elderly population; share of the population with a certain education level; or percentage of migrants) and thus their studies would benefit from applying CoDa techniques.

## 2. Theoretical background

The main idea Aitchison proposed for analysis CoDa data was to transform the data in a way that allows them to be analysed with standard statistical tools, designed for unconstrained data. He therefore introduced the concept of log-ratio transformations: additive log-ratio transformation (alr) and centered log-ratio transformation (clr). In 2003, (Egozcue *et al.*, 2003) proposed a new family of transformations called isometric log-ratio transformations (ilr) to overcome some of the limitations of the alr and clr transformations. However, there is no single best transformation and all of them have their strengths and limitations (McKinley *et al.*, 2020). Furthermore, special attention should be taken in case of zeros, since the logarithm of 0 is undefined (Martín-Fernández, Barceló-Vidal and Pawlowsky-Glahn, 2003; Lloyd, Pawlowsky-Glahn and Egozcue, 2012).

The alr transformation is defined as:

$$\text{alr}(x) = \log\left(\frac{x_1}{x_D}, \frac{x_2}{x_D}, \ldots, \frac{x_{D-1}}{x_D}\right)$$

where x = [$x_1, x_2, ..., x_D$] is the vector of composition components that sum to a constant (Lloyd, Pawlowsky-Glahn and Egozcue, 2012). The alr transformation is useful for parametric modelling. However, it is not invariant under permutation of the components and it is not isometric between the simplex (the sample space of compositional data; x = [$x_1 ... x_D$] ∈ $S^D$) and the real space ($R^{D-1}$) (Aitchison, 2002; Buccianti and Grunsky, 2014). Therefore, the clr transformation was proposed:

$$\text{clr}(x) = \log\left(\frac{x_1}{g(x)}, \frac{x_2}{g(x)}, ..., \frac{x_D}{g(x)}\right)$$

where g(x) is the geometric mean of the part of the composition. The clr transformation solves the problem of symmetry and, unlike alr-transformed variables, ordinary distances can be computed (Lloyd, Pawlowsky-Glahn and Egozcue, 2012). The clr transformation is useful for generating biplots (Aitchison and Greenacre, 2002; Lloyd, Pawlowsky-Glahn and Egozcue, 2012), but it cannot be used for parametric modelling (Buccianti and Grunsky, 2014). Furthermore, both alr and clr transformations may be difficult to interpret. Therefore, (Egozcue *et al.*, 2003) proposed the ilr transformation:

$$y = \text{ilr}(x) = (y_1, y_2, ..., y_{D-1}) \in \mathbb{R}^{D-1}$$

Where

$$y_i = \frac{1}{\sqrt{i(i+1)}} \ln\left(\frac{\prod_{j=1}^{i} x_j}{(x_i + 1)^i}\right) \text{ for } i = 1, ..., D-1$$

The ilr transformation allows using all the standard multivariate procedures (Egozcue *et al.*, 2003). However, ilr coordinates may also be difficult to interpret and the Sequential Binary Partitions method was therefore developed (Egozcue and Pawlowsky-Glahn, 2005; Pawlowsky-Glahn and Egozcue, 2011). The result is a particular case of ilr coordinates (i.e., balances) that represent the relationship between two groups of parts (allowing interpretation within and between groups of parts). The difficulty is to select the correct partitions for obtaining meaningful interpretations and it should be done base on expert knowledge and/or by compositional biplots (Lloyd, Pawlowsky-Glahn and Egozcue, 2012). The general formula for balances is:

$$b_i = \sqrt{\frac{rs}{r+s}} \ln\left(\frac{(\prod_+ x_j)^{\frac{1}{r}}}{(\prod_- x_k)^{\frac{1}{s}}}\right) \text{ for } i = 1, ..., D-1$$

Where $\prod_+$ and $\prod_-$ are the parts coded as + or − in the partitioning scheme and r and s the number of components in the + and − partition. In our case, for example, with three components ($x_1, x_2, x_3$) a partition could be the one shown in Table 1.

Table 1: Example of a partitioning scheme for a three-part component

| x1 | x2 | x3 | Balance (b) | r | s |
|----|----|----|-------------|---|---|
| 1  | 1  | -1 | 1           | 2 | 1 |
| 1  | -1 | 0  | 2           | 1 | 1 |

The two balances wound therefore be:

$$b_1 = \sqrt{\frac{2}{3}}\ln\left(\frac{(x_1 x_2)^{\frac{1}{2}}}{x_3}\right); \quad b_2 = \sqrt{\frac{1}{2}}\ln\left(\frac{x_1}{x_2}\right)$$

## 3. Data and methods

This section introduces the population data as well as the data on housing prices used for the study. Moreover, it describes the application of CoDa techniques on the data.

### 3.1. Population data at parish level

Data at parish level have been obtained from Statistics Denmark (Statistics Denmark, 2021). The table contains the population at the first day of the year and distinguishes between five ancestry groups: persons of Danish origin, immigrants from Western countries, immigrants from non-Western countries, descendants from Western countries, and descendants from non-Western countries. We selected only the data from the Capital region of Denmark in 2020. We also assumed that immigrants and their descendants behave similarly and thus we merged them to simplify the interpretability of our case study. Finally, we closed the dataset to represent percentages over the total population in each parish. The summary statistics of the percentages are shown in Table 2 and the spatial distribution in Figure 1.

Table 2: Summary statistics of population data (in percentage) by parish (N = 127)

|        | Danes       | Non-Western | Western    |
|--------|-------------|-------------|------------|
| Mean   | 77.8        | 14.5        | 7.7        |
| Median | 79.6        | 11.5        | 7.1        |
| IQR    | 72.9 - 84.3 | 7.7 - 17.7  | 4.9 - 9.7  |
| Range  | 21.2 - 94.6 | 2.7 - 69.9  | 2.7 - 19.9 |

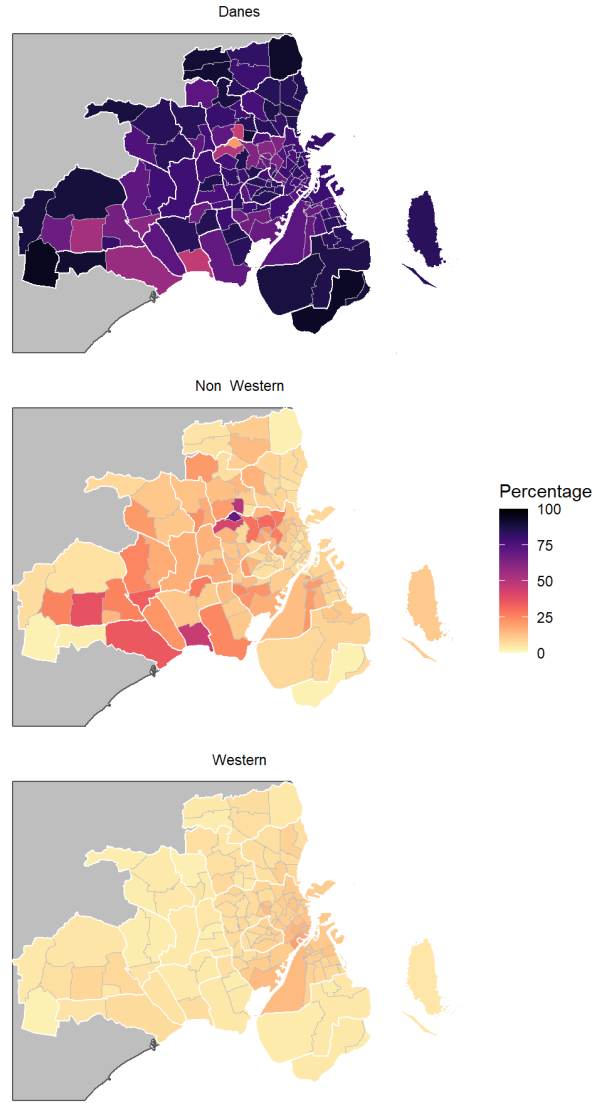

Figure 1: Population distribution [%] in the Capital region of Denmark

### 3.2. House prices

We obtained the individual house prices from the Building and Dwelling Register (BBR - https://teknik.bbr.dk/forside). We used all residences for year-round living (i.e., excluding summer houses and similar seasonal housing) and we selected from the main residential buildings only those that are on the ordinary free trade (sales between parties who are not members of the same family and sales that are not considered as a partial gift) or public sales, assuming that they also represent a market value. Furthermore, we filtered out those dwellings that are not actually used for residential purpose. Dwellings smaller than 10 m$^2$ were also removed from the analysis. Colleges and residential buildings for institutions (i.e., different kinds of dormitories) were excluded from the data analysis since they are mainly outside of the free marked. We also removed properties with no value and calculated a price in 1000

Danish Crowns per square meter (kDKK m$^{-2}$). Finally, we calculated the number of dwellings per parish, as well as the mean and median prices per square meter (Table 3 and Figure 2).

Table 3: Summary statistics at parish level (N = 125)

| Value | N. houses | House prices (kDKK · m$^{-2}$) | |
|---|---|---|---|
| | | Mean | Median |
| Mean | 150.1 | 82.8 | 46.2 |
| Median | 127.0 | 42.3 | 38.0 |
| IQR | 85.0 - 182.0 | 33.2 - 61.2 | 31.1 - 47.4 |
| Range | 7.0 - 637.0 | 22.1 - 1,132.9 | 0.4 - 602.6 |

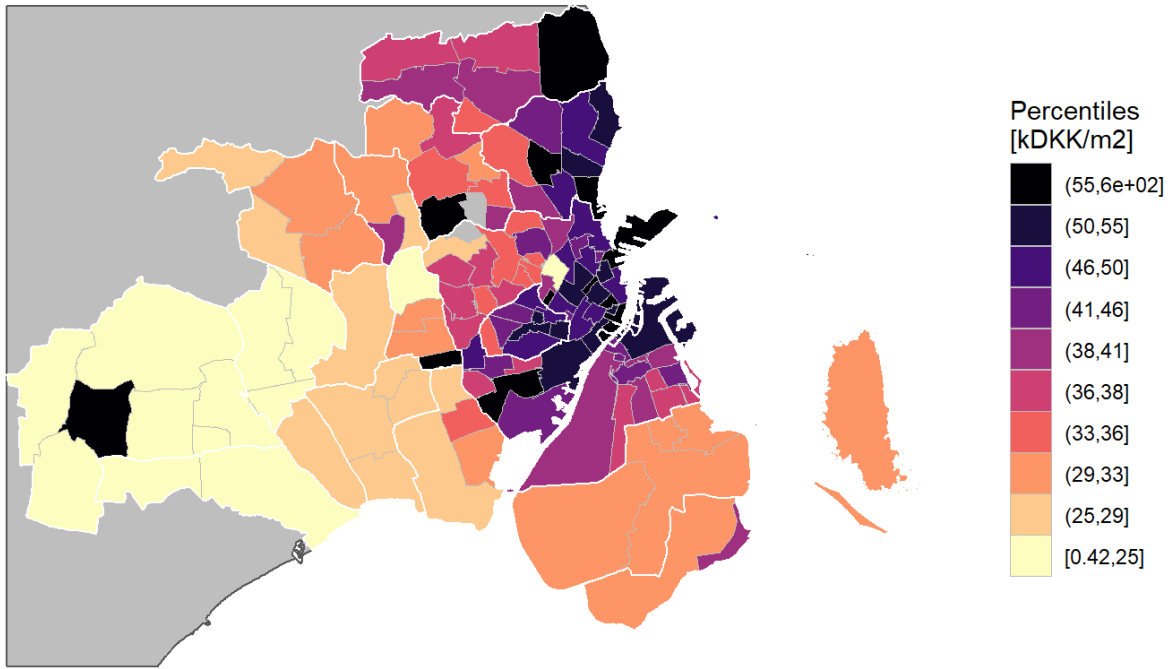

Figure 2: Spatial distribution of parish-level median housing price in 2020

### 3.3. Compositional data

The partition for the balance calculation has been carried out based on the compositional biplot (Figure 3A). The first component differentiates mainly non-Western migrants from other inhabitants, and it accounts for about 73% of the variance (Table 4). The second component explains the remaining 27% and mainly separate Danes population from Western migrants (Table 4).

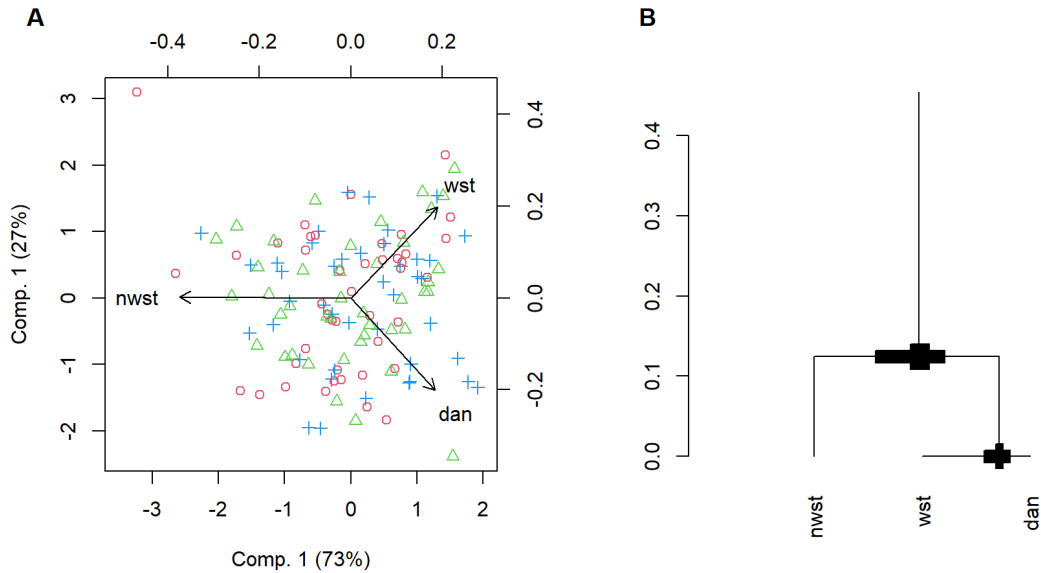

Figure 3: A) Compositional Biplot, and B) balance-dendrogram of the selected partition

Table 4: Importance of components

|  | Comp.1 | Comp.2 |
|---|---|---|
| Standard deviation | 0.574 | 0.353 |
| Proportion of Variance | 0.726 | 0.274 |
| Cumulative Proportion | 0.726 | 1.000 |

We therefore selected the following partition for the balances (Table 5 and Figure 3B), and the equations for estimating the two balances are:

Table 5: Partition scheme

| Danes | Western | Non-Western | Balance | r | s |
|---|---|---|---|---|---|
| 1 | 1 | -1 | b1 | 2 | 1 |
| 1 | -1 | 0 | b2 | 1 | 1 |

$$b_1 = \sqrt{\frac{2}{3}} \ln\left(\frac{(\text{Danes} \cdot \text{Western})^{0.5}}{\text{NonWester}}\right)$$

$$b_2 = \sqrt{\frac{1}{2}} \ln\left(\frac{\text{Danes}}{\text{Western}}\right)$$

Where Danes, Western, and Non-Western are the percentages of each population group in the parish.

Using these balances, we performed a hierarchical cluster analysis to investigate whether there are parishes with similar population distributions and whether they are spatially aggregated or not. We used an agglomerative clustering with the Ward's method, which minimizes the total within-cluster variance. The analysis was carried out with the function hclust (see complementary material) of the R-software (R Core Team, 2021). We also evaluated the spatial autocorrelation of balances, and we carried out a basic linear model to evaluate the association between migration and housing prices. We used the median housing price as dependent variable and the two balances as the independent variables.

$$\ln(HP_i) = \beta_0 + \beta_1 \cdot b_{1i} + \beta_2 \cdot b_{2i} + \varepsilon_i$$

Where $HP_i$ is the median housing price in the parish i; $b_{1i}$ and $b_{2i}$ are the balances for each parish, $\beta = (\beta_0, \beta_1, \beta_2)$ the regression coefficients to be estimated based on the data, and $\varepsilon_i$ is the error term. Contrary to standard regression models, we cannot interpret the coefficients (β) as the increase/decrease of Y due to an increase/decrease of X; instead we need to think in the relative behaviour of the components (Chastin *et al.*, 2015). Furthermore, only the coefficient of the balance that explain the ratio between one component and the others is interpretable (Chastin *et al.*, 2015). In our case (model 1), it is the coefficient for the balance that explains the behaviour of non-Western migrants in relation to Danes and Western migrants (i.e., b1). For interpreting the relative influence of other components, subsequent equivalent models with other components playing the role of the pivotal first component would be needed (Chastin *et al.*, 2015; Dumuid *et al.*, 2018, 2021). We therefore create two more models; i) model 2 for evaluating the relative influence of Western population in relation to Danes and non-Western migrants, and ii) model 3 for the influence of Danes vs. non-Western and Western migrant population.

## 4. Results

This section summarizes the main findings we have obtained in our case study. Our aim was to demonstrate the applicability of CoDa techniques in population geography, not only in migration studies, showing how log-ratio transformations should be used in common (spatial) data analysis techniques (e.g., cluster analysis and regression analysis).

### 4.1. Ternary diagram of population structure

Population structure varies significantly across the capital region of Denmark, as we can observe in the ternary plot centred over the compositional mean (80.1%, 7.4%, 12.5% of Danes, Western, and non-Western population, respectively; Figure 4). Brown colours indicate the parishes with a higher proportion of Danes than the compositional mean, while green colours indicate a higher proportion of Western population and pink colours a higher proportion of non-Western population. As the map shows, Western populations prevail in parishes close to the city centre while non-Western citizens tend to settle down in the Western peripheral parishes with percentages up to 41.6%, 49.61%, and 69.85% for Husumvold, Haralds and Tingbjerg parishes, respectively (all of them clustered in the central part of the figure – pink colours).

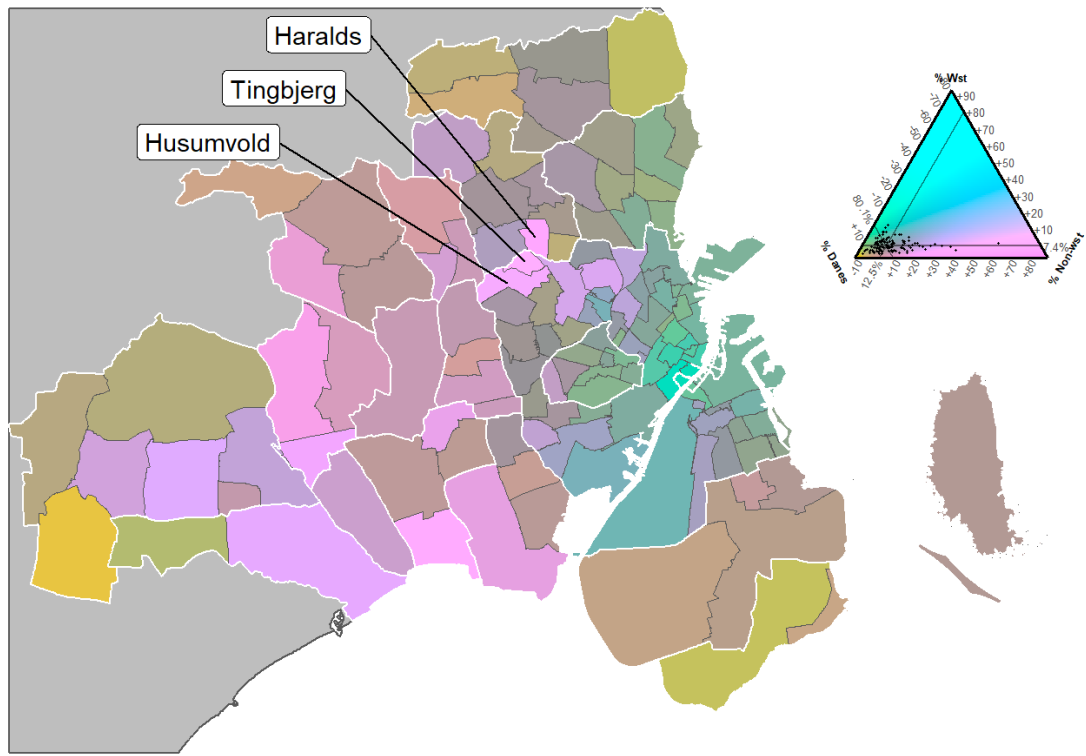

Figure 4: Ternary diagram of the population distribution in 2020 (Danes – people of Danish origin, Wst – Western population, Non-wst – Non-Western population).

## 4.2. Balances

The compositional biplot (Figure 3A) shows the same pattern observed in the ternary diagram. Non-Western migrants dominate component 1, with an opposite direction than Danes and Western migrants. Therefore, we selected a partitioning scheme (Table 5 and Figure 3) that separates mainly non-Western migrants from Danes and Westers migrants (b1), and then Western migrants and Danes (b2). High values of b1 indicate a smaller proportion of non-Western population and high values of b2 a smaller proportion of Western citizens (Figure 5).

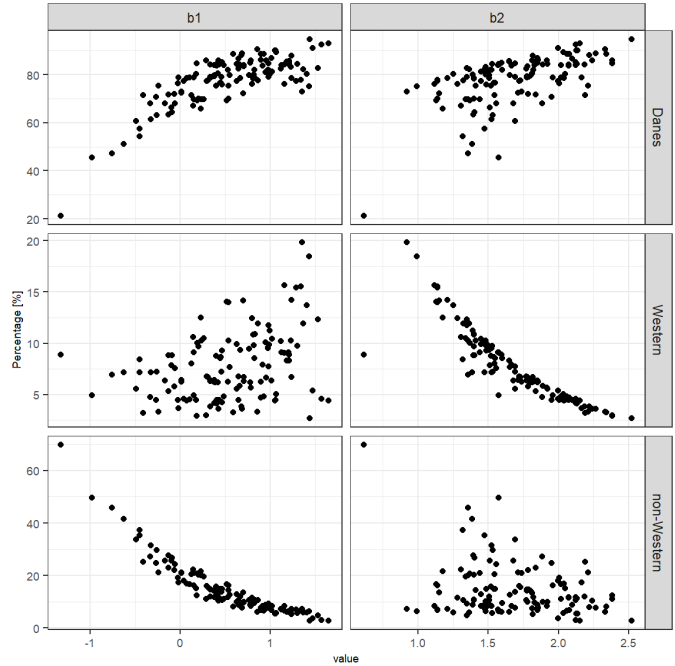

Figure 5: Balances vs. population distribution percentages

The balances show a positive spatial autocorrelation (Table 6), suggesting some degree of spatial structure in the population by its origin. The local indicators of spatial association (LISA - (Anselin, 2010) for both balances (Figure 6) confirm that non-Western migrants tend to live in the peripheral Western parishes (blue colours in b1) while Western migrants tend to settle down around the city centre (blue colours in b2). Furthermore, non-Western migrants try to avoid the Eastern coast of the capital region (red colours in b1). Danes, on the other hand, try to avoid the city centre and the parishes to the south, north and west tend to have high percentage of national residents (red colours in b2).

Table 6: Moran's I for each balance

| Balance | Index | Expectation | Variance | Statistic | P value |
|---|---|---|---|---|---|
| b1 | 0.470 | -0.008 | 0.003 | 8.563 | $5.503 \times 10^{-18}$ |
| b2 | 0.542 | -0.008 | 0.003 | 9.837 | $3.915 \times 10^{-23}$ |

Method: Moran I test under randomisation. Alternative: greater

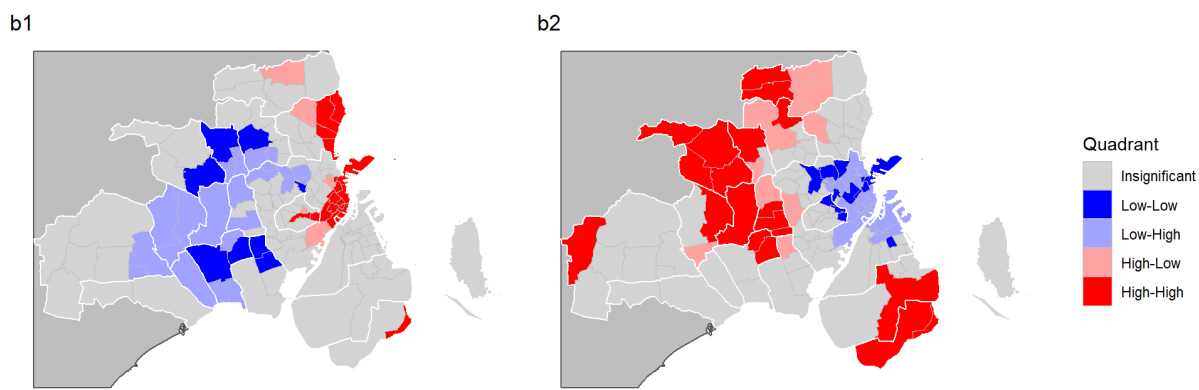

Figure 6: LISA plots of the two balances (b1 and b2)

### 4.3. Hierarchical clustering

We can identify two main clusters in the data (Figure 7, blue lines), with a different proportion of Western migrants. Cluster CL1-2 (blueish colours in Figure 8) has a median percentage of around 10% (Table 7), while the proportion of Western migrants in CL3-4 (orangish colours in Figure 8) is approximately 5% (Table 7). This supports the findings we saw before, where the Western migrant population tends to live in central parishes. Then, these two main clusters are further divided into four clusters (Figure 7, orange lines). The difference is mainly due to an increase of non-Wester population. In this regard, CL2 has a higher proportion of non-Western population than CL1 (i.e. 24.1% and 7.8%, respectively; Table 7), while CL4 has a higher concentration than CL3 (i.e. 20.5% and 8.9%, respectively; Table 7). Again, we observe a preference of peripheral Western parishes for non-Western population (CL2 - light blue; and CL4 – light orange; Figure 8). Finally, CL3 shows the parishes with the highest proportion of national citizens, with values around 85.9% (Table 7).

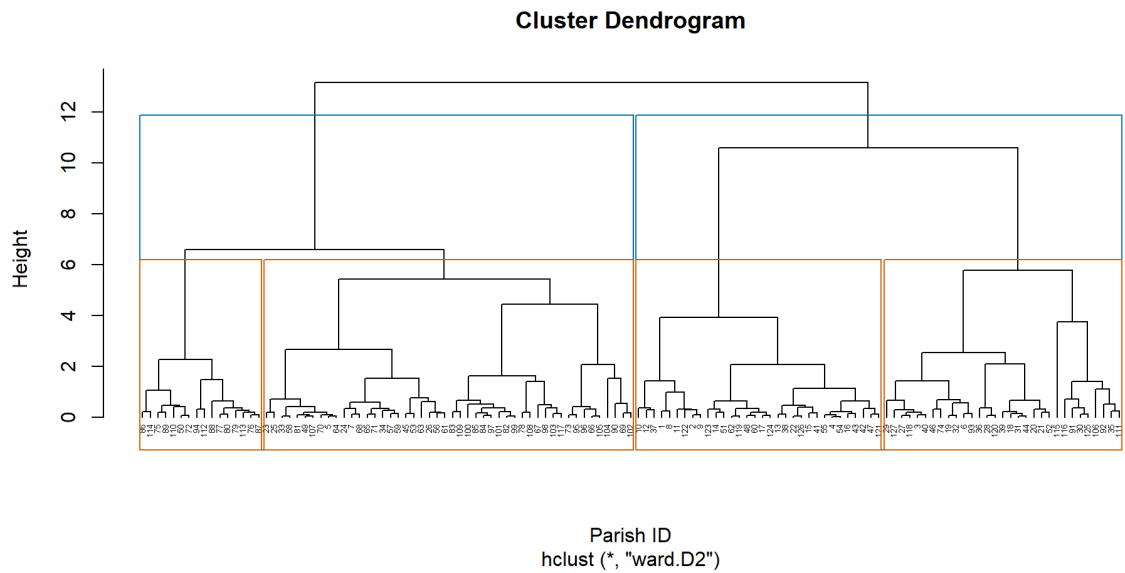

Figure 7: Cluster dendrogram with the two balances

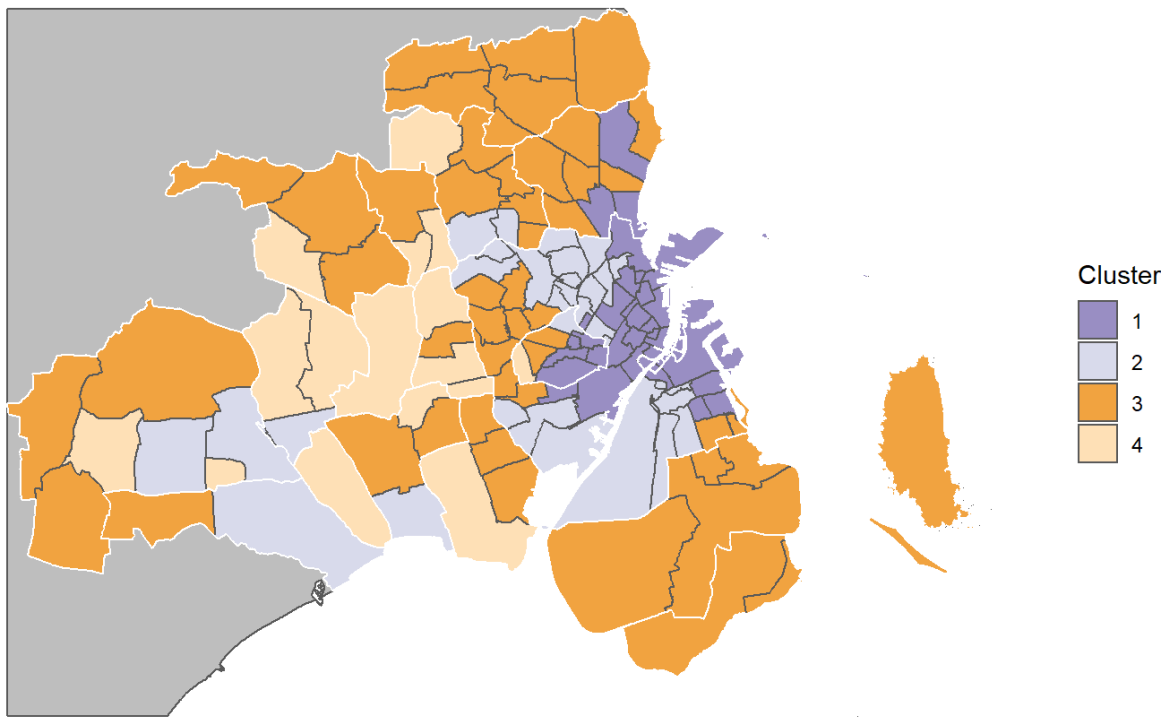

Figure 8: Spatial distribution of the four clusters

Table 7: Compositional mean of the four clusters

| Cluster | N Parishes | Danes | Western | Non-Western |
|---------|------------|-------|---------|-------------|
| 1 | 32 | 81.0 | 11.2 | 7.8 |
| 2 | 31 | 66.6 | 9.3 | 24.1 |
| 3 | 48 | 85.9 | 5.2 | 8.9 |
| 4 | 16 | 74.8 | 4.7 | 20.5 |

### 4.4. Linear model

The pivotal first component of the models are shown in Table 8. The coefficient (β) for the balance that pivot around non-Western vs. Danes/Western population (i.e., $b_1$) is highly significant and positive (Model 1; β = 0.359; p-value < 0.001) indicating that the median housing price increases with a decrease of non-Western migrants (and an increase in Danes and Western population). We also see that a decrease on Western population (increase of Danes and non-Western) is associated with a decrease of median prices (Model 2; β = -0.430; p-value = 0.002). Finally, we do not observe an association between house prices and the relative variation of Danes in relation to foreign population (Model 3; β = 0.072; p-value = 0.622).

Table 8: Regression coefficient (β) of pivot coordinates

| model | Estimate | CI (lower) | CI (upper) | Std. Error | t value | Pr(>|t|) | |
|-------|----------|------------|------------|------------|---------|----------|---|
| *Model 1 (dan & wst vs. nwst)* | 0.359 | 0.166 | 0.551 | 0.097 | 3.683 | <0.001 | *** |
| *Model 2 (dan & nwst vs. wst)* | -0.430 | -0.699 | -0.162 | 0.136 | -3.170 | 0.002 | ** |
| *Model 3 (nwst & wst vs. dan)* | 0.072 | -0.217 | 0.360 | 0.146 | 0.494 | 0.622 | |

Note: dan = Danes; nwst = non-Western population; wst = Western population: CI = 95% confidence interval

### 5. Discussion

The hieratical clusters analysis over the log-ratio transformed data has allowed us to detect four main spatial clusters, which clearly discriminate parishes according to their population structure (i.e., people of Danish origin, Western migrants, and non-Western migrants). As expected, Danes are the main population in the region. However, they tend to avoid the city centre and the parishes to the south, north and west of the city centre tend to have high percentage of national residents. Western migrants, on the other hand, prefer the central areas while the composition of non-Western migrants increase in the Western peripheral parishes. Additionally, the ternary diagram (*Figure 4*) has also allowed us to graphically identify parishes with a very high percentage of non-Western migrants; i.e., Husumvold, Haralds and Tingbjerg parishes with values up to 41.6%, 49.61%, and 69.85%; respectively, and manly as a cost of Danes (51.20%, 45.47%, 21.24%; respectively) rather than Western migrants (7.20%, 4.91%, 8.90%; respectively). These parishes are examples of what the Danish authorities call parallel societies or "ghettos", which trigger political actions (Gulis, Safi and Linde, 2020; Seemann, 2020). Our observations also are in agreement with previous studies (Georgati and Keßler, 2021) and confirm the spatial population segregation in the Capital region of Denmark.

It is important to note that different phenomena can lead to the same proportions in the data. Compositions only give information about the relative magnitude of its components but not the absolute values (Aitchison, 2002), and additional information would be needed in order to make inferences with the absolute values. In our study, for example, we have seen the spatial segregation of the population by its origin but we cannot interpret its causes; e.g., if it is due to a socio-economic (status) segregation or that immigrants tend to settle down in areas with an ethnicity background similar to their country of origin. This is actually a limitation of any observational study, which helps to make hypothesis about the phenomena we are investigating but further studies would be needed to verify them. CoDa techniques are however more robust than ordinary statistics methods because they alleviate issues with spurious correlations and they avoid problems with subcompositions, since results obtained using the whole dataset do not contradict results obtained with only a subset (Aitchison, 2002).

Analysing the housing prices in the four clusters, there are some clear differences in the mean and median values (Table 9). In general, CL1 and CL2 have higher values (i.e. mean around 57.0 and 41.7 kDKK m$^{-2}$; and median of 50.7 and 38.6 kDKK m$^{-2}$, respectively) than CL3 and CL4 (i.e. means of 38 and 27 kDKK m$^{-2}$, and medians of 35.6 and 27.0 kDKK m$^{-2}$, respectively). We interpret that this difference is probably more related to location than population structure. On the one hand, CL1 and CL2 parishes are central parishes close to the city centre where we can expect more amenities and, therefore, people may be willing to pay more for their house in these areas than for further away ones. On the other hand, the central area of the capital region is more densely populated than the periphery (Georgati and Keßler, 2021) so we can expect a higher demand for houses in these parishes which may lead to an increase in the housing prices. However, if we compare the values within the two main clusters (i.e., CL1 vs. CL2 and CL3 vs. CL4), the prices are lower in the parishes where the proportion of non-Western population is relatively high (around 20%, Table 7).

Table 9: House price statistics (kDKK·m$^{-2}$) in each cluster

| Cluster | 1 (N = 32) | 2 (N = 29) | 3 (N = 48) | 4 (N = 16) |
|---|---|---|---|---|
| **Mean values (kDKK·m$^{-2}$)** | | | | |
| Mean | 119.8 | 82.9 | 65.7 | 59.9 |
| Median | 57.0 | 41.7 | 38.2 | 27.2 |
| IQR | 50.7 - 98.7 | 37.3 - 60.8 | 31.4 - 49.8 | 24.8 - 33.5 |
| Range | 39.4 - 1,132.9 | 22.1 - 554.6 | 22.2 - 554.7 | 22.5 - 333.6 |
| **Median values (kDKK·m$^{-2}$)** | | | | |
| Mean | 67.6 | 37.3 | 35.8 | 50.9 |
| Median | 50.7 | 38.6 | 35.6 | 27.0 |
| IQR | 46.9 - 54.5 | 34.2 - 42.0 | 30.1 - 40.2 | 24.7 - 32.0 |
| Range | 37.1 - 602.6 | 0.4 - 70.8 | 20.5 - 55.6 | 5.1 - 367.3 |

The ternary diagram with the median housing prices and the population distribution (Figure 9) also shows the differences in the median prices by parishes and their association with the clusters. The parishes of CL1, where the proportion of Western migrants is the highest and the proportion of non-Western migrants is relatively low, has the highest medina housing prices. In CL2, the median price decreases with the increase of the non-Western migrants. In these parishes the proportion of non-Western migrants change hugely, from around 10% to up to approximately 45% (there are two parishes with even more proportion of non-Western migrants; i.e. Haralds and Tingbjerg with 49.61%, and 69.85% respectively, but there were no data of housing sales on the ordinary free trade or public sales in these parishes in 2020).

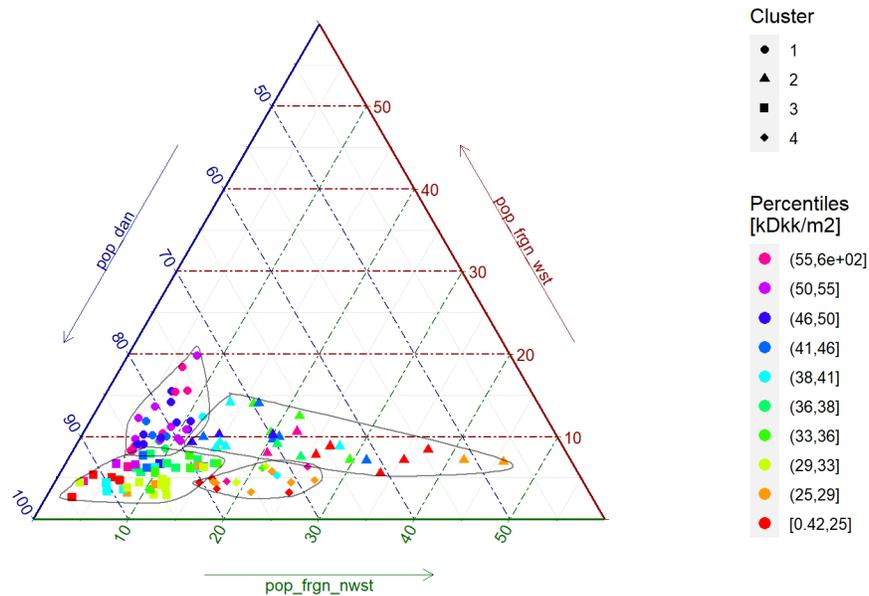

Figure 9: Population distribution (in percentage) and housing prices (medina values in kDKK m$^{-2}$)

The relative numbers of non-Western and Western migrants are statistically significantly associated with the median housing prices (Table 8). However, it does not mean that they are independent predictors of the median prices by themselves and their interpretation should be done in comparison with the other variables. A useful tool that helps to interpret the overall influence of the variables on the predicted mean is to plot the outcome of the model under different compositions; both in ternary diagrams and bivariate plots with the confidence interval (Figure 10). Our result suggests, as we saw in Table 8, that there is a negative association between housing prices and the increase of non-Western migrants in relation to Danes and Western populations; i.e., green-red colours in the diagram towards an increase of the percentage of non-Western migrants. Furthermore, the blue-pink colours for the increase of Western migrants show the positive association between this group (in relation to the other two) and housing prices.

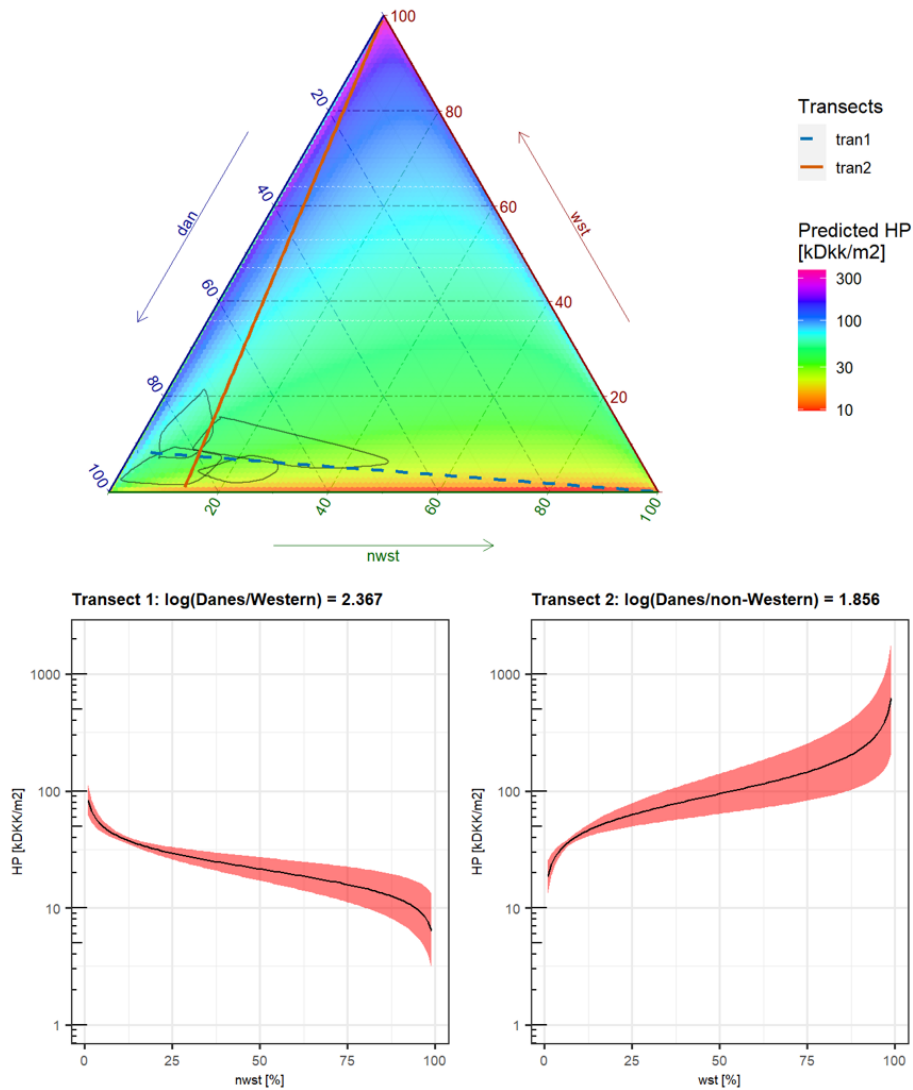

Figure 10: Ternary diagram showing the predicted median housing price at parish level for different compositions (top), and predicted median house prices at parish level with 95% CI for transects 1 and 2 (bottom). The circles represent the clusters – see Figure 9

While our analysis is limited to a small number of socio-economic variables, the results point to two different kinds of segregation that can be observed in the Danish capital region. The clustering of native Danes and Western immigrants on the one hand and high concentrations of non-Western immigrants in a relatively small number of parishes on the other hand indicates racial segregation. Taking into account house prices, we further see that Western migrants are concentrated in the central parishes, which are characterized by higher prices, whereas the parishes with high numbers of non-Western show significantly lower house prices. This correlation therefore also indicates socio-economic segregation.

# 6. Conclusions

CoDa techniques are more robust and appropriate than standard statistical and geostatistical methods when we are analysing close data (e.g. percentages) because they avoid spurious correlations, predictions outside the range, and have no problems with sub-compositional coherence. However, they still are not widely used in population geography. Therefore, we have carried out a case study to evaluate their applicability in this field to evaluate how we should interpret the results obtained with CoDa techniques. In this regard, we have analysed the population distribution, divided into people of Danish origin, Western immigrants, and non-Western immigrants in the Capital region of Denmark and its possible relationship with housing prices.

Our analysis helps to interpret the spatial population segregation in the Capital region of Denmark. We have detected four main cluster regions with clearly different population structure in terms of migration percentages. Furthermore, we have found that migration patterns might help to interpret some of the variation in the median housing prices at parish level. However, we showed the potential of CoDa analysis in regression models but we did not try to make an exhaustive evaluation of the association between them. In this sense, we did not take into account many other factors that can affect housing prices. Our results, therefore, have to been taken with caution and further investigations should be carried out to evaluate the causal relationship between migration and housing prices in the Capital region of Denmark.

We show how balances can be used for alleviating the issue of data interpretation with CoDa methods. There is still some complexity in the interpretation of models based on balances, and they can been seen, somehow, as black-boxes (especially when we have more than three components in our dataset). However, our exercise is a good example of how regression models should be interpreted when we have compositional data as explanatory variables. Our study is a basic case study with only three components but it can be generalized to other population datasets.

**Acknowledgements**

This work has been supported by the European Union's Horizon 2020 research and innovation programme under grant agreement No 870649, the project Future Migration Scenarios for Europe (FUME; https://futuremigration.eu); and by the Aalborg University Strategic Fund, through the project: "Global flows of migrants and their impact on north European welfare states – FLOW" (www.flow.aau.dk).

**Disclosure statement**

No potential conflict of interest was reported by the author.

**Data availability**

The data and R-scripts that support the findings of this study are openly available in https://github.com/javiereliomedina/CoDa_migr_cph.git.